\documentclass[traditabstract]{aa} % for the abstract without structuration 
 \usepackage{longtable}
\usepackage{epsfig}
\usepackage{graphicx}
\usepackage{xcolor}
\usepackage{txfonts}
\usepackage{amssymb}
\usepackage{natbib}                          
\bibpunct{(}{)}{;}{a}{}{,}
\newcommand{\be}{\begin{equation}}
\newcommand{\ee}{\end{equation}}
\newcommand{\bea}{\begin{eqnarray}}
\newcommand{\eea}{\end{eqnarray}}

\begin{document}

% \title{Accretion flow hydrodynamics in intermediate polars for dipole geometry.}
% \title{Interaction between accretion disc and magnetosphere in intermediate polar GK Per.}
\title{GK Per and EX Hya: Intermediate polars with small magnetospheres}
\author{
V. Suleimanov\inst{1},
V. Doroshenko\inst{1},
L. Ducci\inst{1,2},
G.V. Zhukov\inst{3},
K. Werner\inst{1}}

\offprints{V.~Suleimanov}
\mail{e-mail: suleimanov@astro.uni-tuebingen.de}

\institute{
Institut f\"ur Astronomie und Astrophysik, Kepler Center for Astro and
Particle Physics,
Universit\"at T\"ubingen, Sand 1,
 72076 T\"ubingen, Germany
\and
{ISDC Data Center for Astrophysics, Universit\'e de Gen\`eve, 16 chemin d'\'Ecogia, 1290 Versoix, Switzerland}
\and
Kazan (Volga region) Federal University, Kremlevskaja str., 18, Kazan 420008, Russia
}

\date{Received xxx / Accepted xxx}

   \authorrunning{Suleimanov et al.}
   \titlerunning{GK Per and EX Hya: small magnetospheres}

\abstract
{Observed hard X-ray spectra of intermediate polars are determined mainly by
the accretion flow velocity at the white dwarf surface, which is normally close
to the free-fall velocity. This allows to estimate the white dwarf masses as
the white dwarf mass-radius relation $M-R$ and the expected free-fall
velocities at the surface are well known. This method is widely used, however,
derived white dwarf masses $M$ can be systematically underestimated because the
accretion flow is stopped at and re-accelerates from the magnetospheric
boundary $R_{\rm m}$, and therefore, its velocity at the surface will be lower
than free-fall.
To avoid this problem we computed a two-parameter set of model hard X-ray spectra, which allows to constrain a
degenerate $M - R_{\rm m}$ dependence. On the other hand, previous works showed
that power spectra of accreting X-ray pulsars and intermediate polars exhibit
breaks at the frequencies corresponding to the Keplerian frequencies at
the magnetospheric boundary. Therefore, the break frequency $\nu_{\rm b}$ in an intermediate
polar power spectrum gives another relation in the $M -
R_{\rm m}$ plane. The intersection of the two dependences allows, therefore, to determine
simultaneously the white dwarf mass and the magnetospheric radius. To verify the method we analyzed the archival
\emph{Suzaku} observation of EX Hya obtaining $M /M_\odot= 0.73 \pm 0.06$ and $R_{\rm m}/R =
2.6\pm0.4$  consistent with the values determined by other authors. Subsequently, we
applied the same method to a recent \emph{NuSTAR}
observation of another intermediate polar GK~Per performed during an outburst and found
$M/M_\odot= 0.86 \pm 0.02 $ and $R_{\rm m}/R = 2.8\pm0.2$.
The long duration observations of GK Per in
quiescence performed by \emph{Swift}/BAT and INTEGRAL observatories indicate
increase of magnetosphere radius $R_{\rm m}$ at lower accretion rates.}


\maketitle
%
%________________________________________________________________

\section{Introduction}
Close binary systems consisting of a normal donor and a white dwarf (WD)
accreting through Roche lobe overflow are named cataclysmic variables
\citep[CVs, see review in ][]{Warn:03}. Intermediate polars (IPs) are a
subclass of CVs with moderately magnetized WDs ($B \sim 10^4 - 10^6$~G). Central parts
of accretion discs in these systems are destroyed by the WD magnetic field 
within its magnetosphere with radius $R_{\rm m}$. At smaller radii
the accreted plasma couples to the field lines and forms a shock wave close to the WD
magnetic poles. The shocked plasma is heated up to WD virial temperatures ($kT
\sim 10 - 30$~keV), cools through  thermal bremsstrahlung, and settles down to
the WD surface. As a result, IPs are bright hard X-ray sources
\citep{Revnivtsev.etal:04a, Barlow.etal:06, Landi.etal:09}. The hot post-shock
region is optically thin and its averaged temperature can be estimated directly
from the observed spectrum. The plasma temperature depends only on WD compactness, and therefore provides a direct estimate of the WD mass
\citep{Rothschild.etal:81}.

A theory of the post-shock regions (PSR) on WDs was first considered by
\citet{Aizu:73} and further developed in several works where the WD gravity,
the influence of cyclotron cooling, the dipole geometry of the magnetic field, and the
difference between the temperatures of the electron-ion plasma components were taken
into account \citep{FPR:76, Wu.etal:94, Cropper.etal:99,Canalle.etal:05,
Saxton.etal:07, HI:14a}. Some of these models were used to estimate the WD
masses in several intermediate polars and polars \citep{Cropper.etal:98,
Cropper.etal:99, Ramsay:00, revnivtsev.etal:04b, SRR:05, Falanga.etal:05,
Betal:09, Yuasa.etal:10, HI:14b}. However, in all these models the accreting
plasma was assumed to fall from  infinity whereas in reality the
magnetospheric radius could be small enough (a few WD radii) to break this
assumption. As a consequence the accretion flow will be accelerated to lower
velocities and the post-shock region will have lower temperature for the given
WD parameters. 
Therefore, WD masses derived from previous PSR models
can be underestimated.
% previously:
%On the other hand, for a given temperature this will lead to
%underestimated WD mass.

Such possibility has been first suggested by \cite{SRR:05} for GK
Per to explain a significant discrepancy
between the WD mass estimated using RXTE observations of an outburst and optical spectroscopy.
Later \citet{Betal:09} confirmed this conclusion by estimating the WD mass in
GK~Per in quiescence. The difference between the estimates of the WD mass in EX
Hya obtained using the RXTE observations \citep[0.5 $M_\odot$, ][]{SRR:05} and
from the optical observations \citep[0.79 $M_\odot$, ][]{BR:08} was also
attributed to a small magnetospheric radius in this system \citep[$\sim$ 2.7 WD
radii, see e.g.][]{Revnivtsev.etal:11}.

Here we present a new set of IP model spectra which for the first time
quantitatively account for the finite size of the magnetosphere. The models are
calculated for a set of magnetospheric radii (expressed in units of WD radii)
assuming a relatively high local mass accretion rate ($>$\,1
g\,s$^{-1}$\,cm$^{-2}$). Additional cyclotron cooling was ignored as it is not
important in such conditions. The model is publicly available and allows direct
investigations of the dependence of magnetospheric radius on mass accretion
rate in IPs.

A new method of simultaneous determination of the WD mass and the
magnetospheric radius is suggested on the base of this set. 
We propose to add the information about the observed frequency of a break 
in power spectra of the IP X-ray light curves.
Using the spectral fitting and this additional information we can
constrain the relative magnetospheric radius together with the WD mass. We
verified the method using high quality \emph{Suzaku} observations of the well
studied IP EX Hya, and subsequently applied it to study the spectra of the
IP GK Per in outburst and quiescence.
  
\section{The method}

% In this section we present the idea of a new method for simultaneous
% determination of the WD mass and the magnetospheric radius $R_{\rm m}$ using
% hard X-ray observations of intermediate polars.
The basic commonly accepted physical picture of the X-ray emitting region in
intermediate polars is that the matter falls along magnetic field lines onto the
WD and forms an adiabatic shock above the surface. The free-fall velocity
of the matter ${\rm v}_{\rm ff}$ decreases by a factor of four as it crosses the
shock according to the Rankine-Hugoniot relations. The rest kinetic energy
transforms to internal gas energy and heats the matter up to the temperature

\be \label{kt0}
     kT_0 = \frac{3}{16}\mu m_{\rm H}\,{\rm v}_{\rm ff}^2 = \frac{3}{8}\mu m_{\rm H}\,\frac{GM}{R}\left(1-{r_{\rm m}}^{-1}\right).
\ee
Here $M$ and $R$ are the mass and the radius of the WD, $r_{\rm m} = R_{\rm
m}/R$ is the relative radius of the magnetosphere, $\mu=0.607$ 
is the mean molecular weight for a completely ionized plasma with solar chemical composition,
and $m_{\rm H}$ and $k$ are the proton mass and the Boltzmann constant. Here we take
into account that the matter starts falling from the finite distance $R_{\rm
m}$ from the WD. The heated matter settles down to the WD surface in the sub-sonic
regime and loses energy by optically thin bremsstrahlung. The cooling due to
cyclotron emission is insignificant in intermediate polars due to a relatively
weak WD magnetic field and can be ignored.
The height of the shock wave above the WD surface is determined by the cooling rate
and models of the post-shock region can be accurately computed together
with emergent spectra (see next Section).

\begin{figure}
\centering
\includegraphics[angle=0,scale=1.1]{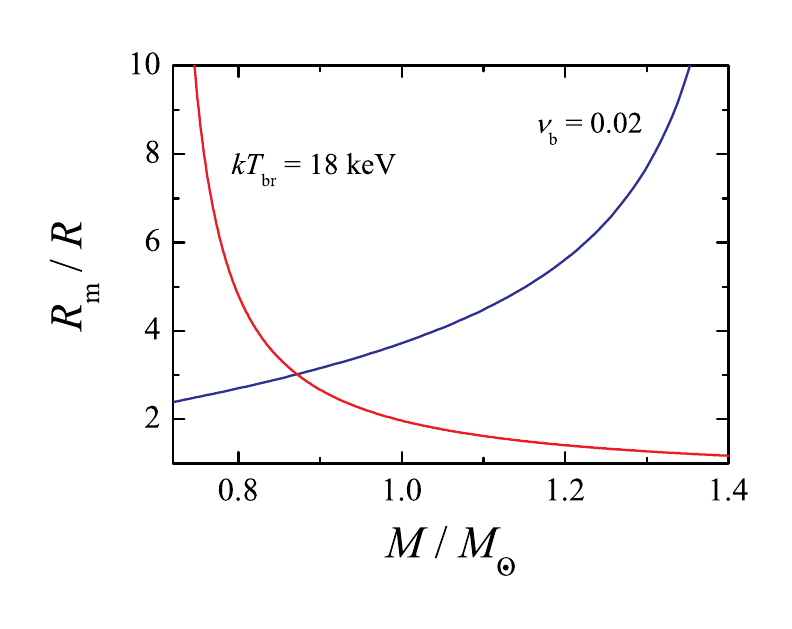}
\caption{\label{fig1a}
Curves of the constant bremsstrahlung temperature (\ref{mwd}) and the constant break frequency (\ref{rrm}) on the $m - r_{\rm m}$ plane.
Intersection of the curves provides an estimate of the WD mass and the magnetospheric radius. 
}
\end{figure}

\begin{figure}
\centering
\includegraphics[angle=0,scale=0.8]{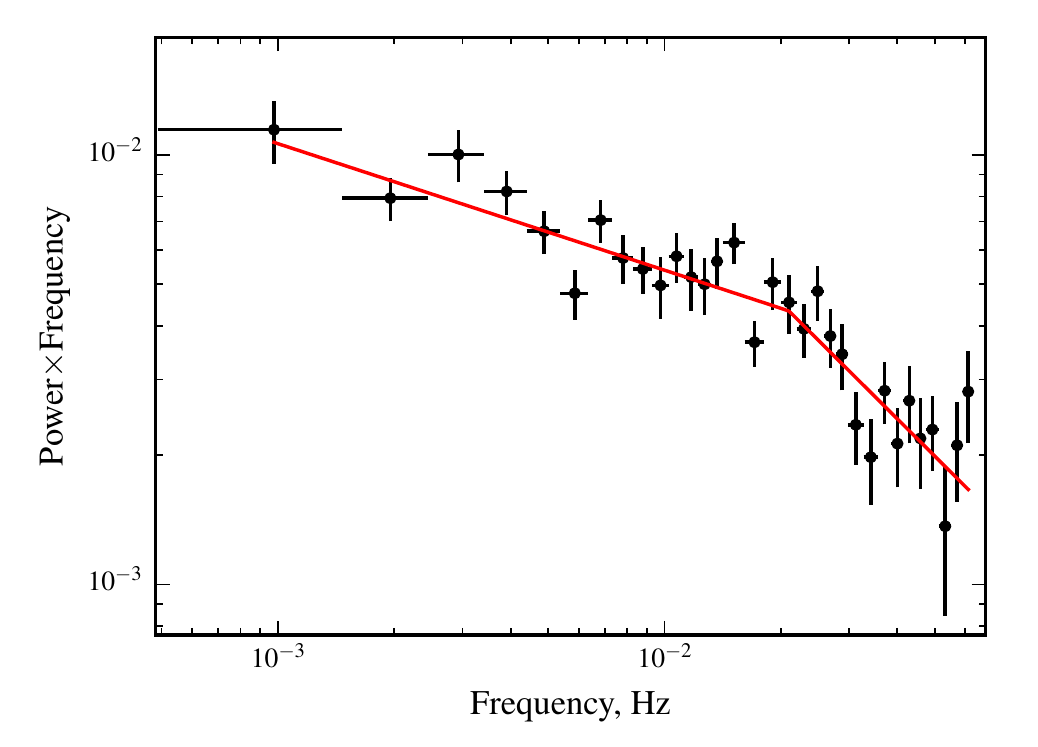}
\caption{\label{fig1b}
Power spectrum of the 0.5-10\,keV X-ray light curve of the EX Hya observed by \emph{Suzaku}. The break frequency $\nu_{\rm b} = 0.021\pm 0.006$
is consistent with value reported by Revnivtsev et al. (2011) using RXTE data.  
}
\end{figure}

We note that the observed hard X-ray spectra of intermediate polars can be well
approximated with thermal bremsstrahlung. The temperature of the
bremsstrahlung $kT_{\rm br}$ is, however, lower in comparison with $kT_0$, and a
proportional factor $A$ between these temperatures can be found from 
accurate PSR computations only. The computations and their comparison with the
observed X-ray spectra presented later give
\be \label{ktbr}
       kT_{\rm br} = A\,kT_0 \approx 0.64\,kT_0.
\ee 
The WD radius depends on the WD mass, see, e.g. \citet{Nbg:72}:
\be \label{rwd}
         R = 7.8\times 10^8\,{\rm cm}\,\left(\left(\frac{1.44}{m}\right)^{2/3}-\left(\frac{m}{1.44}\right)^{2/3}\right)^{1/2},
\ee
where $m = M/M_\odot$ is the WD mass in units of solar masses. Therefore, Eqs.
(\ref{kt0} - \ref{rwd}) define a curve in the $m - r_{\rm m}$ plane,
which corresponds to the observed bremsstrahlung temperature $kT_{\rm br}$
(see. Fig.\,\ref{fig1a}).
Another curve on the $m - r_{\rm m}$ plane can be obtained from X-ray light curve
analysis. \citet{Revnivtsev.etal:09, Revnivtsev.etal:11} showed that power
spectra of X-ray pulsars and intermediate polars exhibit a break at the frequency
$\nu_{\rm b}$ corresponding to the Kepler frequency at $R_{\rm m}$:
\be \label{nub}
        \nu_{\rm b}^2 = \frac{GM}{4\pi^2\,R_{\rm m}^3}.
\ee
An example of the power spectrum with the break obtained for the intermediate
polar EX Hya using the observations performed by \emph{Suzaku} observatory 
(see Sect. \ref{subsect EX Hya}) is presented in Fig.\,\ref{fig1b}. 
The curve described by Eq.\,(\ref{nub}) 
is also shown in Fig.\,\ref{fig1a}. 
The intersection of the two curves
%%The two curves intersect at some $m$ and $r_{\rm m}$ which 
allows to estimate the WD mass and the
magnetospheric radius of the investigated intermediate polar.

The Eqs.\,(\ref{kt0} - \ref{rwd}) and (\ref{nub}) can be simplified using
the linear approximation of the WD mass-radius relation presented by
\citet{SPFW:08}
\be \label{rwdl}
  R \approx 1.364\times  10^9\,{\rm cm}\, (1-0.59\,m),
\ee
which is valid for WD masses in the 0.4-1.2 $M_\odot$ range. The resulting
relations are
\be \label{mwd}
         m \approx \frac{kT_{\rm br}}{15\,(1-r_{\rm m}^{-1})+0.59\,kT_{\rm br}},
\ee
where $kT_{\rm br}$ is measured in keV, and
\be \label{rrm}
         r_{\rm m} \approx 1.5 \,\left(\frac{\nu_{\rm b}}{0.02}\right)^{-2/3}\,\frac{m^{1/3}}{1-0.59\,m}. 
\ee

The accuracy of these relations can be checked using available data. The hard
X-ray spectrum of the  intermediate polar V1223 Sgr was fitted by a bremsstrahlung
spectrum with $kT_{\rm br} \approx 29$\,keV \citep{Revnivtsev.etal:04a}. We get
$m \approx$\,0.9 using relation (\ref{mwd}) with $r_{\rm m} = \infty$, whereas
the direct fitting with PSR spectra gave 0.95$\pm$0.05 \citep{SRR:05}.
\citet{Revnivtsev.etal:11} found $r_{\rm m} \approx 2.7$ for EX Hya using fixed
$m=0.79$ and $\nu_{\rm b} = 0.02$. The relation (\ref{rrm}) gives $r_{\rm m}
\approx$\,2.6.

Thus, relations (\ref{mwd}) and (\ref{rrm}) allow to evaluate the WD mass
and the magnetospheric radius using a simple X-ray spectrum fitted by
bremsstrahlung and  break frequency in the power spectrum. However, direct
fitting of PSR model spectra is necessary for more accurate results. Of course,
we use the accurate $M-R$ relation further on and provide the relations above
only for convenience.

\section{The model of the post-shock region}
\label{sect. The model of post-shock region}

\subsection{Basic equations}

%Below we derive the equations for the post-shock accretion column (PSAC) in the 
%dipole geometry.
The post-shock
region can be fully described using the following set of equations \citep[see, for example,][]{Mih78}. The continuity equation is
\be \label{cont}
%      \frac{\partial \rho}{\partial t} +
       {\bf \nabla \cdot} (\rho {\bf v})
=0,
\ee
where $\rho$ is the plasma density, and $\bf v$ is the vector of the gas
velocity. Conservation of momentum for each gas element is described by the
vector Euler equation
\be \label{euler}
%\rho \frac{\partial {\bf v}}{\partial t} + 
  \rho {\bf v \cdot \nabla v} = -{\bf \nabla}P +{\bf f},
\ee
where $P$ is the gas pressure (we ignore the radiation pressure here), and ${\bf
f}$ is the force density. The energy equation for the gas is
%\begin{eqnarray} 
\be
\label{energy}
%\frac{\partial}{\partial t} \left( \frac{1}{2} \rho {\rm v}^2 +\varepsilon\right) + 
 {\bf \nabla \cdot}
\left[ \left(\frac{1}{2} \rho {\rm v}^2 + \varepsilon + P\right){\bf v}\right] =
%\\ \nonumber 
{\bf f \cdot v - \nabla \cdot q}.
\ee
%\end{eqnarray}
Here $\varepsilon = (3/2) P$ is the density of internal gas energy. The first
term on the right side of the energy equation is the power density, 
the second term accounts for radiative energy loss (${\bf
q}$ is the vector of the radiation flux). These equations must be supplemented by
the ideal-gas law
\be \label{igl}
   P =  nkT =\frac{\rho k T}{\mu m_{\rm H}},
\ee 
where $n$ is the total number density of particles. 

Here we consider the one-dimension optically thin flow in a white dwarf
gravitational field. Therefore, we substitute the radiation loss term by the local
radiative cooling function $\Lambda(\rho, T)$
\be \label{cf}
   {\bf \nabla \cdot q} = \Lambda(\rho,T) = n_{\rm i} n_{\rm e} \Lambda_{\rm N}(T),
\ee
where 
\be
n_{\rm e} = \frac{\rho}{\mu_{\rm e} m_{\rm H}}
\ee
 is the electron number density, and 
 \be
 n_{\rm i} = n-n_{\rm e} = \frac{\rho} {m_{\rm H} } \left(\frac{1}{\mu} - \frac{1}{\mu_{\rm e}}\right) 
 \ee
is the ion number density. Here $\mu_{\rm e} = 1.167$ is the mean number of
nucleons per electron for fully ionized solar mix plasma. The universal
cooling function was computed by many authors and in previous work
\citep{SRR:05} we used $\Lambda_{\rm N}$ computed by
\citet{SD:93}. In the present work we use a more modern cooling function computed by
the code {\sl APEC} \citep{APEC} using the database {\sl
AtomDB}\footnote{http://www.atomdb.org} for a solar chemical composition.
In the previous work \citep{SRR:05} the total cooling function was
overestimated by factor $\approx 4$, as it was correctly mentioned by
\citet{HI:14a}, because $n^2$ instead of $n_{\rm i}\, n_{\rm e}$ was used in
Eq.\,\ref{cf}. Fortunately, this error did not influence the emergent
model spectra and the obtained WD masses (see Appendix). It was corrected in 
subsequent work \citep{SPFW:08}.
  
We assume that only gravity force operates in the PSR considered here 
\be
{\bf f} = -\frac{GM}{(R+z)^2}\,\rho = -g(z)\,\rho.
\ee
Here $z$ is geometrical height above the WD surface, and the considered accretion flow settles along
$z$.    

\subsection{Quasi-dipole geometry}

We use the approximation of the dipole geometry suggested by \citet{HI:14a}.
They assumed that the PSR cross-section $S$ depends on $z$ as follows: $S \sim z^n$. 
In this case every divergence in hydrodynamical equations can be written
as ${\bf \nabla \cdot y} = S^{-1} d(Sy)/dz$:
\be \label{su1}
  \frac{1}{S}\,\frac{d(S\rho \rm v)}{dz} = 0,
\ee    
\be
    \rho{\rm v}\,\frac{d{\rm v}}{dz} = -\frac{dP}{dz} - g(z)\,\rho,
\ee
and
\be
        \frac{1}{S}\,\frac{d}{dz}\left[S{\rm v}\left(\frac{1}{2} \rho {\rm v}^2 + \varepsilon + P\right)\right] =
         -g(z)\,\rho\,{\rm v} -\Lambda(\rho,T).
\ee
If we take $S \sim z^2$, we will obtain the well known equations for the spherically symmetric geometry. The dependence
$S \sim z^3$ mimics the dipole geometry.

Eq.\,(\ref{su1}) has the integral
\be \label{cont_int}
   S\rho\,{\rm v} = a,
\ee 
where $a$ is the local mass accretion rate at the WD surface, $[a] =$\,g\,s$^{-1}$\,cm$^{-2}$.
Using this integral we can replace $\rho$ in the next two Eqs. by $a/S{\rm v}$ and finally we have
\be \label{eiler}
     {\rm v}\frac{d{\rm v}}{dz} + \frac{S{\rm v}}{a}\,\frac{dP}{dz} = -g(z),
\ee 
and
\be \label{energ}
 {\rm v}\frac{dP}{dz} +\gamma\,P\,\frac{d{\rm v}}{dz} +\gamma\frac{P{\rm v}}{S}\,\frac{dS}{dz} =-(\gamma -1)\Lambda(\rho,T),
\ee
where $\gamma=5/3$ is the adiabatic index. We note that Eq.\,(\ref{eiler})
coincides with the corresponding equation in \citet{HI:14a}, with the exception
of the energy conservation law. However, our Eq.\,(\ref{energ}) coincides with
the energy equation in \citet{Canalle.etal:05} written for the PSR at the
magnetic pole (in this case $\frac{1}{h_2}\,\frac{d}{d\omega} = \frac{d}{dz}$),
if we take into account that their product $h_1h_3$ corresponds to our function
$S$.

Equations  (\ref{eiler}) and (\ref{energ}) can be rewritten as follows
\be \label{eiler1}
      \frac{d{\rm v}}{dz} =  \frac{(\gamma-1)\,S\,\Lambda(\rho,T)+  a\,g(z)- \gamma P\,{\rm v} \,dS/dz} {S\gamma\,P - a{\rm v}}
\ee 
and
\be \label{energ1}
 \frac{dP}{dz} = -g(z)\frac{a}{S{\rm v}} -\frac{a}{S}\,\frac{d{\rm v}}{dz},
\ee
where
\be
         S= \left(\frac{R+z}{R}\right)^3,
\ee
and
\be
       \frac{dS}{dz} =3\,\frac{(R+z)^2}{R^3}.
\ee 

\subsection{Method of solution}

Equations (\ref{eiler1}) and (\ref{energ1}) can be solved with appropriate
boundary conditions. We use a commonly accepted suggestion about a strong
adiabatic shock. In particular, at the upper PSR boundary ($z=z_0$) we have
\be \label{bct}
       {\rm v_0}= -\frac{1}{4} {\rm v_{\rm ff}}(z_0),~~~~~~P_0 = -3\frac{a{\rm v_0}}{S(z_0)},~~~~~T_0= 3\frac{\mu m_{\rm H}}{k}\,{\rm v}_0^2
\ee  
and
\be \label{bcb}
      {\rm v} = 0
\ee
at the WD surface ($z=0$). The free-fall velocity at the upper PSR boundary
${\rm v_{\rm ff}(z_0)}$ depends on the WD compactness and the inner disc radius,
or the magnetosphere radius $R_{\rm m}$
\be \label{vff}
    {\rm v_{\rm ff}}(z_0) = \sqrt{2GM}\,\left(\frac{1}{R+z_0} - \frac{1}{R_{\rm m}}\right)^{1/2}.
\ee 
Therefore, the input parameters for each PSR model are the WD mass $M$,
the local mass accretion rate at the WD surface $a$, and the magnetospheric
radius $R_{\rm m}$. 
We reduced the number of parameters of the model
by eliminating the WD radius using the WD mass-radius relation of \citet{Nbg:72}.

\begin{figure}
\centering
\includegraphics[angle=0,scale=1.]{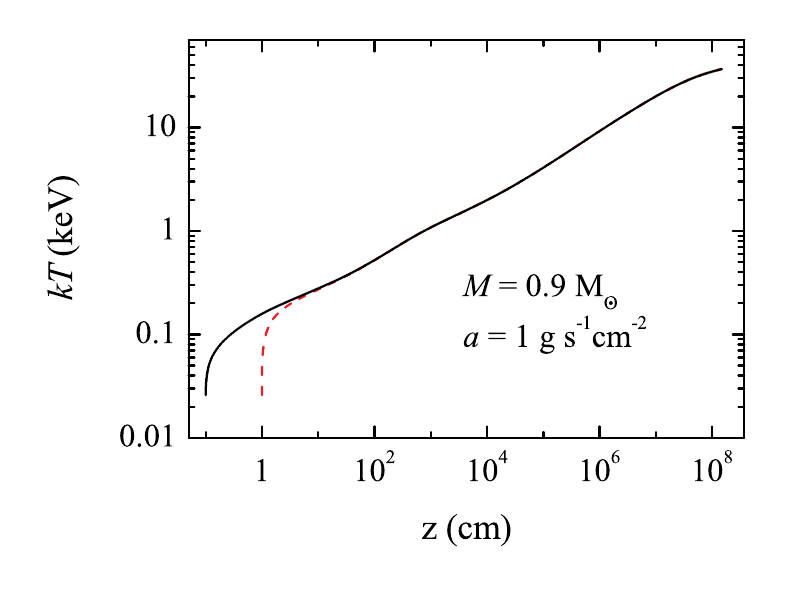}
\caption{\label{fig1}
Comparison of the temperature structures of the models computed for two different values of the first model point, 
$z_1 = 0.1$~cm, solid curve, and $z_1=1$~cm, dashed curve. 
}
\end{figure}

We computed each PSR model using a logarithmically equidistant grid over $z$.
We start from the first point with very small $z_1=0.1$~cm. We fix the
temperature $T_1 = 300\, 000$~K and some pressure $P_1$ at this point. The
corresponding density $\rho_1$ and velocity ${\rm v}_1$ are found using the
mass conservation law (\ref{cont_int}) and the equation of the state
(\ref{igl}). Then Eqs. (\ref{eiler1}) and (\ref{energ1}) are integrated up to
the height $z$, where ${\rm v}(z) = 0.25\,{\rm v}_{\rm ff}(z)$. At that point
the pressure has to be also equal to $-3a{\rm v}(z)/S(z)$. However, the last
condition does not hold for the arbitrarily chosen $P_1$. Therefore, we find
the required pressure at the first point $P_1$ in the range 0.1-30\, $P_0$ by
the dichotomy method. Once both aforementioned upper boundary conditions are
satisfied with a relative accuracy better than $10^{-8}$, we take the height
$z$ as the PSR height $z_0$. The results depend slightly on the choice of the
starting point (see Fig.\,\ref{fig1}). In the previous paper \citep{SRR:05}
models were computed starting from the highest PSR point $z_0$. Both
methods give coincident solutions at $kT > 0.5$~keV, but the new one describes
the bottom part of the PSR much better.
 
\begin{figure}
\centering
\includegraphics[angle=0,scale=1.]{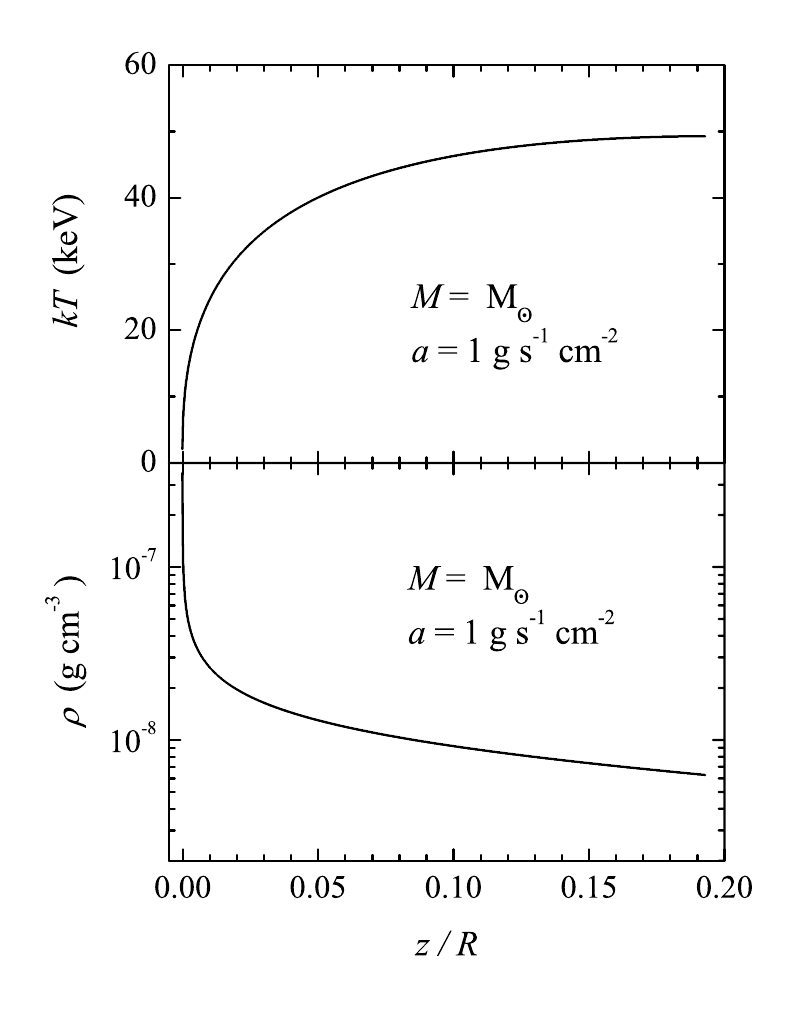}
\caption{\label{fig2}
Distributions of  temperature and density in the model with $M =
M_{\odot}$ and  local mass accretion rate $a= 1$~g s$^{-1}$ cm$^{-2}$. See
the similar Fig.2 in \citet{HI:14a}.
}
\end{figure}

\begin{figure}
\centering
\includegraphics[angle=0,scale=1.]{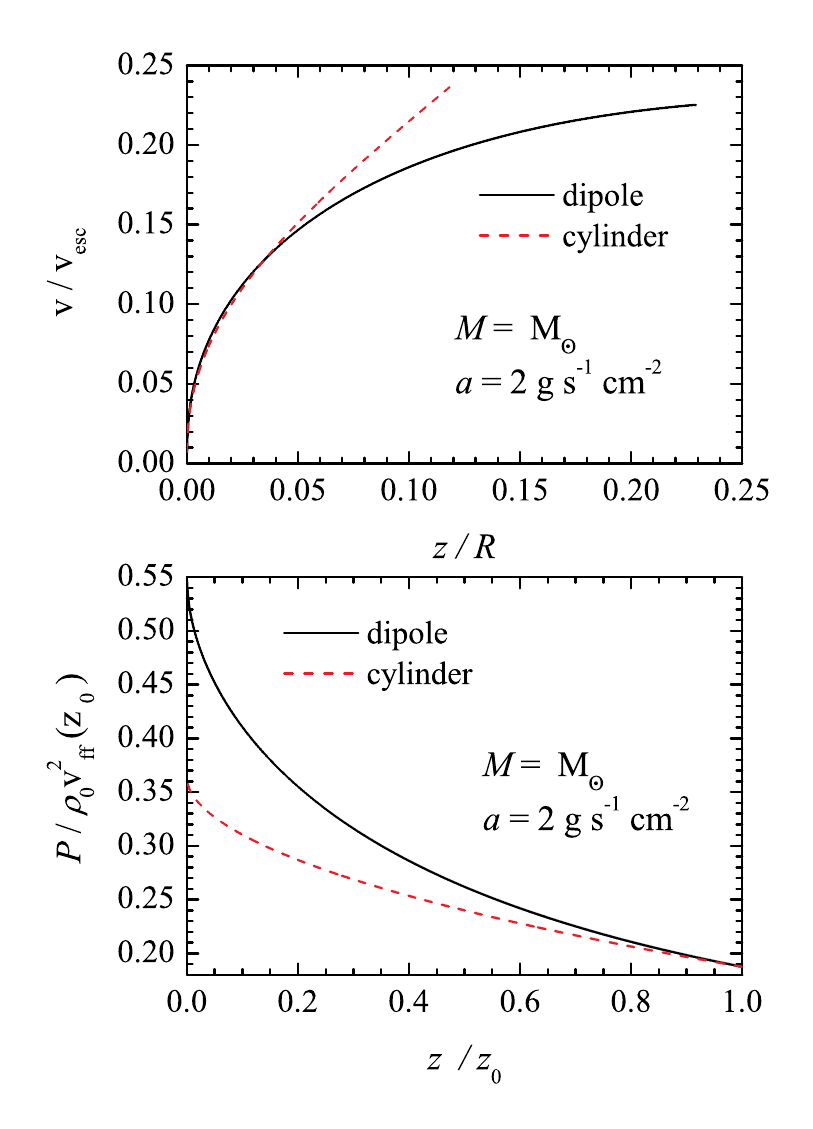}
\caption{\label{fig3}
Comparison of the models computed in cylindrical and dipole geometries. 
This result reproduces Figs.\,6 and 4 in the paper by \citet{Canalle.etal:05}. 
Here ${\rm v_ {esc}} = \sqrt{2GM/R}$.
}
\end{figure}

The code allows to compute models in the cylindrical geometry, too. The
computation of a such model with the same parameters,
as were used by \citet{HI:14a} for one of the models, gives similar results
\citep[compare Fig.\,\ref{fig2} with figure\,2 in][]{HI:14a}. 
We also find a good agreement between our results and those
obtained by \citet{Canalle.etal:05} for the dipole geometry 
\citep[compare Fig.\,\ref{fig3} with figure\,6 in][]{Canalle.etal:05}

We note, however, that the results obtained by \citet{HI:14a} for the dipole
geometry are physically incorrect. They found that the PSR heights are lower in
the dipole geometry. 
However, in the dipole geometry scenario, the plasma
cools slowly because of its lower density compared to the cylindrical geometry case.
Therefore, the PSR height in the dipole geometry is expected to be higher.

\subsection{Spectra computations}
We computed the PSR model spectra assuming solar abundances and
and fully ionized of all the abundant elements  in the PSR. 
This assumption is reasonable at temperatures 
$kT >$~1 keV, therefore, the computed spectra are correct at 
relatively high energies only ($E > 3$\, keV). At lower energies,
the spectra are dominated by numerous emission spectral lines and
photo-recombination continua \citep[see, e.g.][]{Canalle.etal:05}. 
%We assumed
%solar mix of the chemical elements in all the considered here models. 
The PSR models are optically thin, and the relative spectra can be calculated by simple
integration of the local (at the given height $z$) emissivity coefficients $\eta_{\rm E} = k_{\rm
E}\,B_{\rm E}$
% \textcolor{red}{\emph{maybe should explain $B_E$, $k_E$}} 
and taking into account the dipole geometry
\be \label{sp}
       F_{\rm E} = \int_0^{z_0} k_{\rm E}\,B_{\rm E}\,S\,dz,
\ee
where $B_{\rm E}$ is Planck function.
The free-free opacities $k_{\rm E}$ for the 15 most abundant chemical elements were
computed using Kurucz's code ATLAS \citep{Kurucz:70, Kurucz:93} modified for
high temperatures \cite[see details in][]{Ibragimov.etal:03, sw:07}.

\begin{figure}
\centering
\includegraphics[angle=0,scale=1.]{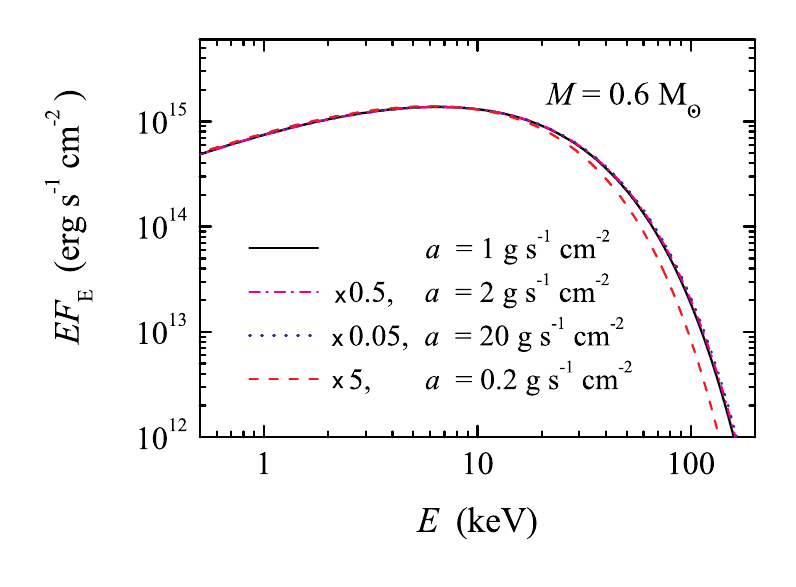}
\caption{\label{fig4}
Comparison of  model spectra for several local mass accretion rates and a fixed
WD mass. Spectra are normalized to the spectrum with $a$~ = 1 g s$^{-1}$ cm$^{-2}$.
}
\end{figure}

\begin{figure}
\centering
\includegraphics[angle=0,scale=1.]{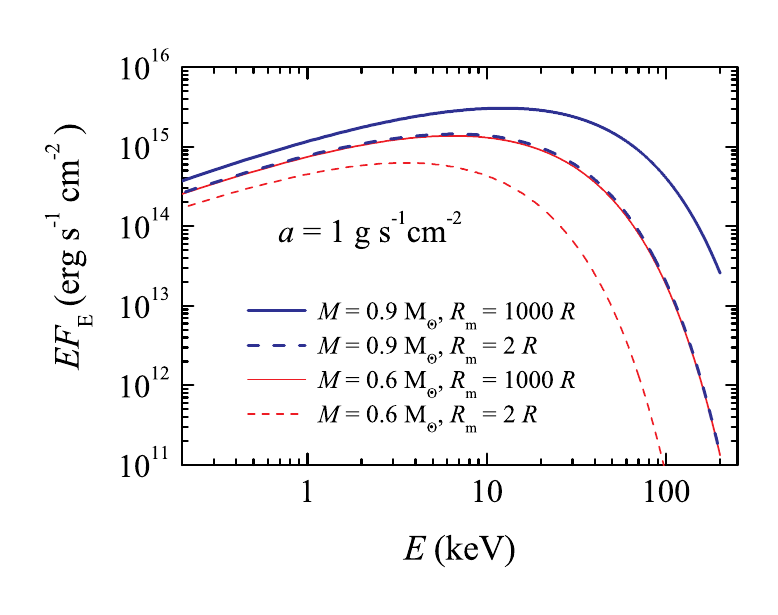}
\caption{\label{fig5}
Comparison of  model spectra computed for two WD masses, 0.6 (thin curves)
and 0.9 (thick curves) $M_{\odot}$ and two relative magnetospheric radii,
$R_{\rm m}/R$~= 2 (dashed curves) and 1000 (solid curves). The local mass
accretion rate is fixed to $a=1$\,g\,s$^{-1}$\,cm$^{-2}$ for all the
computed spectral models.
}
\end{figure}

Examples of model spectra are presented in Figs.\,\ref{fig4} and
\ref{fig5}. The first one shows that spectra have a degeneracy with respect to the
local mass accretion rate and are almost indistinguishable for any $a >$~1 g
s$^{-1}$ cm$^{-2}$. The model spectrum computed for $a=0.2$ g s$^{-1}$
cm$^{-2}$ is, however, slightly softer. Therefore, we can compute the model
spectra for every WD mass with one sufficiently high $a$ and it would be
sufficient to evaluate the WD mass by fitting its hard X-ray
spectrum with the computed PSR model spectra. 
On the other hand, different
inner disc or magnetospheric radii are very important and can
influence the WD mass evaluation significantly (Fig.\,\ref{fig5}).

Using the described method we computed a set of PSR model spectra for a grid of
two input parameters, the WD mass $M$ and the relative magnetospheric radius
$R_{\rm m}/R$. { The maximum PSR temperature $T_{0}$ strongly depends on the
factor $1-R/R_{\rm m}$, see Eqs.(\ref{bct}) and (\ref{vff}). Therefore, the
equidistant $T_0$ sub-grid for a given WD mass has to be proportional to 
$R_{\rm m}/R \sim N_{\rm max}/N$.
We chose the sub-grid with $N_{\rm max}=60$, and $N$
changing from 40 (which corresponds to $R_{\rm m}/R=1.5$) to 1 ($R_{\rm
m}/R$=60) . The additional model with $R_{\rm m}/R$=1000 was included to
represent the pseudo-infinity magnetospheric radius. The grid was computed for
56 values of WD mass, from 0.3 to 1.4 $M_{\odot}$ with a step of 0.02
$M_{\odot}$, i.e. 2296 models in total.} Every PSR model was computed for a
fixed mass accretion rate $\dot M = 10^{16}$\, g s$^{-1}$ and  fixed ratio of
the PSR footprint area to the WD surface $f=S_{\rm PSR}/4\pi R^2 = 5\cdot
10^{-4}$. The local mass accretion rate changes from $a \approx 1$~ g s$^{-1}$
cm$^{-2}$ for the lightest WD to $a \approx 70$~ g s$^{-1}$ cm$^{-2}$ for the
heaviest WD in accordance with decreasing WD radius. 
This model grid can be used to estimate
the WD mass at fixed $R_{\rm m}/R$ by fitting an IP hard
X-ray spectrum. In addition, the fit returns the normalization of the spectrum:
\be
      K = \frac{fR^2}{d^2}.
\ee 
The grid will be distributed
as an XSPEC additive table model{\footnote{https://heasarc.gsfc.nasa.gov/xanadu/xspec/newmodels.html}}
and publicly available to the scientific community.

\section{Investigated objects}

\subsection{EX Hya} \label{sect EX Hya}

This intermediate polar is one of the closest CVs with a distance of
$d \approx$\,65 pc and a low accretion luminosity $L_{\rm acc} \approx 2.6 \times
10^{32}$\,erg s$^{-1}$ \citep{Beuermann.etal:03}. The orbital period of EX Hya
is relatively short ($P_{\rm orb}$~= 98.26 min), whereas the spin period is
relatively long ($P_{\rm s}$~= 67.03 min) in comparison with other intermediate
polars. The WD mass in this object is well determined from optical observations ($M / M_\odot = 0.79 \pm 0.03$, 
\citealt{BR:08}). Recently, \citet{Revnivtsev.etal:11} found that the
magnetospheric radius in EX Hya is small, $R_{\rm m}/R \approx 2.7 $. 
A similar value ($R_{\rm m}/R \approx 2.5$) was found before by
\citet{Belle.etal:03}. 
%{\bf 
Moreover, \citet{Sieg.etal:89} found the center of the optical radiation in this system is situated 
at a distance of about $1.5\times 10^9$\,cm from the WD center. Most probably, this light
arises due to irradiation of the inner disc rim with the hard X-ray emission of the PSR.
This distance again corresponds  to $\sim 2 R$ for the $0.8 M_{\odot}$ white dwarf.
Another evidence of the small magnetospheric radius comes from the permanent white dwarf spin period
increasing \citep{Hel:92}. It is possible if the magnetospheric radius is much smaller than the corotation radius, which is indeed a 
very large for EX Hya ($\sim 50 R$).  If the corotation radius is close to the magnetospheric one then the intermediate polar
will be close to the equilibrium with changing spin-up and spin-down periods \citep[see, e.g. the case of FO Aqr in][]{Patterson:98}.  
%}

Such a small magnetospheric radius 
can explain the small WD masses obtained from the fit of the observed spectra
with PSR models (0.5$\pm$0.05 $M_\odot$, \citealt{SRR:05}, and 0.42 $\pm$ 0.02, \citealt{Yuasa.etal:10}) 
when an infinity magnetospheric radius was used in the computations. 
The low absorbed X-ray spectrum of EX Hya 
allows detailed investigation of the spectral lines \citep[see, e.g. ][]{Luna.etal:15}. 
These properties make EX Hya an ideal target for testing the method of
simultaneous determination of the WD mass and the magnetospheric radius proposed in this paper.

\subsection{GK Per} \label{gkper}

The intermediate polar GK Per \citep{Watson.etal:85} 
has a very long
orbital period ($P_{\rm orb} \approx 1.99$~days) and a K1 sub-giant secondary
star \citep{Wrn:76, Crampton.etal:86, MR.etal:02}. GK Per is also
classified as a dwarf nova with very long outbursts ($\approx 50$~days) repeated roughly
every 3 years \citep{Sim:02}. GK Per is also known as Nova Persei 1901 and
based on ejecta studies, \citet{Wrn:76} determined the distance to GK Per of
460 pc which is consistent with a previous determination (470 pc) made by \citet{McL:60}.

The spin period of the WD in GK Per is relatively short \citep[$P_{\rm s} \approx
351$~s, ][]{Watson.etal:85, Mauche:04}. The pulse profile shape changed from a
single-peaked during the outbursts \citep{Hellier.etal:04} to a two-peaked type
in quiescence \citep{Patt:91, Ishida.etal:92}. 
The most likely explanation for these variations is the
obscuration of the second WD pole by a dense accretion curtain during the
outbursts \citep{Hellier.etal:04, Vrielmann.etal:05}. During the outbursts,
quasi-periodic flux oscillations with a typical time-scale
$\sim$~5000 sec were reported both in X-rays \citep{Watson.etal:85} and in the
emission line spectrum in the optical band \citep{MR.etal:99}.

Analysis of the absorption line spectra of the  GK Per secondary during 
quiescence allowed to determine the stellar mass ratio in the system ($q = 0.55
\pm 0.21$) and put lower limits for the binary components, $M/M_{\odot} \ge 0.84
\pm 0.24 $ and $M_{\rm sec} / M_{\odot } \ge 0.48 \pm 0.32$
\citep{MR.etal:02}. These values are in accordance with the masses 
obtained by \citet{Crampton.etal:86}, $M/M_{\odot} \approx 0.9 \pm 0.2 $
and $M_{\rm sec} /M_{\odot}\approx 0.25$. A similar WD mass estimate was
obtained by \citet{Rein:94}, $M/ M_{\odot}\ge 0.78$. 
The large uncertainties in mass determination are mostly 
due to the poorly constrained
inclination of the orbital plane of the system to the line of sight $i$, which
was evaluated to be between 73$\degr$ and $\sim$ 50$\degr$ \citep{Rein:94,
MR.etal:02}. 
\citet{Wrn:86} claimed $i\sim 75\degr$ using the correlation between
the equivalent widths of some emission lines and the inclination angle. This
high inclination angle is preferable from our point of view, because the
optical emission lines (H$_{\beta}$, HeII 4686, and others) are observed in
emission also during the outbursts and no wide absorption wings are observed
\citep{Crampton.etal:86, MR.etal:99}. This behavior is typical for
high-inclined CVs with high mass accretion rates \citep[e.g. UX
UMa,][]{Neustroev.etal:11}. Therefore, the WD mass is likely close to the lower
limit in the allowed region, $M/M_{\odot} \approx 0.9$.

It is also possible to determine the WD masses in old novae by comparing the
observed optical light curves of nova outbursts 
with those predicted by theoretical models. Using this
approach \citet{HK:07} determined for GK Per a WD mass of $M /M_{\odot}= 1.15
\pm 0.05$. We note, however, that this result depends on the assumed
hydrogen mass fraction ($X$=0.54) in the expanded Nova Persei 1901 envelope.

GK Per is a bright hard X-ray source during the outbursts and it was observed
by many X-ray observatories including {\it EXOSAT} \citep{Watson.etal:85,
Norton.etal:88}, {\it Ginga} \citep{Ishida.etal:92}, {\it ASCA} \citep{EI:99},
{\it RXTE} \citep{Hellier.etal:04, SRR:05}, {\it XMM-Newton}
\citep{Vrielmann.etal:05, EH:07}, {\it Chandra} \citep{Mauche:04}, {\it
INTEGRAL} \citep{Barlow.etal:06, Landi.etal:09}, and {\it Swift}\,/\,BAT
\citep{Betal:09}. Some of these observations were also used to estimate
the WD mass.

The first measurements of the WD mass from X-ray data
were based on the cylindrical PSR models \citep{SRR:05} 
and the iron line diagnostic method \citep{EI:99}. 
The estimated WD masses turned out to be 
relatively low (0.59$\pm$0.05 and 0.52 $M_{\odot}$
respectively) and inconsistent with previous estimates based on optical observations (
$\sim 0.9 M_{\odot}$ \citealp{Crampton.etal:86, MR.etal:02}). \citet{SRR:05}
attributed this discrepancy to the fact that the PSR models used to estimate
the mass were computed assuming an accretion flow falling from infinity ($R_{\rm m}
= \infty$ in Eq.\,\ref{vff}) while in reality the co-rotation radius in this
system should be smaller than $\sim10\,R$. Moreover, the
magnetospheric radius has to be even smaller during the outburst, as the
observed X-ray flux increases by more than a magnitude with respect to
quiescence \citep{Ishida.etal:92}. Therefore, the accreting matter is expected
to have lower kinetic energy at the WD surface and to reach lower temperatures
after the shock, so the WD mass will be underestimated if one assumes that the
accretion flow accelerates from infinity.

\citet{Betal:09} used {\it Swift}\,/\,BAT 
observations during the low-flux state of GK\,Per and
took into account the reduction of the kinetic energy. They
obtained an improved WD mass value $0.90\pm0.12\,M_{\odot}$,
consistent with estimates based on optical spectroscopy.
%{\bf
 Nevertheless, recent \emph{Suzaku} observations of GK Per
before and during the latest outburst do not show any significant differences
between the spectrum in quiescence just before the outburst and the spectrum on the middle of the flux rise
\citep{Yuasa.etal:16}. This behaviour of GK Per has to be investigated further.
%}

In the next sections we re-visit this estimate and investigate the dependence of the
magnetospheric radius on X-ray luminosity using the archival {\it
RXTE} and {\it Swift}\,/\,BAT observations and the self-consistent PSR models
presented in Sect. \ref{sect. The model of post-shock region}.

\section{Observations and data analysis} 

\subsection{Suzaku observations of EX Hya} 

\emph{Suzaku} is equipped with the four-module X-ray Imaging Spectrometer (XIS)
\citep{XIS} covering the 0.2-12\,keV energy range, and a collimated Hard X-ray
Detector (HXD) covering the 10-70\,keV and 50-600\,keV energy range with PIN
and GSO detectors \citep{HXD1, HXD2}. \emph{Suzaku} observed EX Hya on Jul 8th
2007 for $\sim100$\,ks (obsid \#402001010) with effective exposure of $\sim
75$\,ks for XIS and 85\,ks for HXD PIN. The source is not detected
significantly in the GSO energy band, so we restricted the analysis to XIS and HXD
PIN data.

For data reduction we follow the standard procedures and employ default
filtering criteria as described in the \emph{Suzaku} data reduction
guide\footnote{https://heasarc.gsfc.nasa.gov/docs/suzaku/analysis/abc/}. We use
the HEASOFT~6.16 software package with the current instrument calibration files
(CALDB version 20151105). For HXD PIN background subtraction we adopted the
``tuned'' non-X-ray background (NXB) event file provided by the HXD team. We
ignored the contribution of cosmic X-ray background (CXB) as it is negligible
compared to the source count rate.

\subsection{Observations of GK Per}

To investigate the properties of the source in outburst we use the recent
$\sim80$\,ks long \emph{NuSTAR} target of opportunity observation performed in
April 2015 (obsid 90001008002). We use the HEASOFT~6.16 software package and
current instrument calibration files (CALDB version 20151105). We follow the
standard procedures described in the instruments data analysis software guide
\footnote{http://heasarc.gsfc.nasa.gov/docs/nustar/analysis/nustar\_swguide.pdf}
to screen the data and to extract the light curves and spectra. Total useful
exposure after screening is $\sim42$\,ks. For timing analysis we combine the
data from the two \emph{NuSTAR} units whereas for spectral analysis we extract
and model spectra from the two units separately.

We note that another observation has been performed with \emph{NuSTAR} in
quiescence (Sep 2015, obsid 30101021002), however, the data are not public.
Therefore, to assess the source properties outside of the outburst we rely on
\emph{Swift}/BAT \citep{BAT} and
INTEGRAL \citep{INTEGRAL} data. The
\emph{Swift}/BAT 70-month survey \citep{BATs} provides
mission long spectra and light curves for all detected sources including GK Per
(listed as SWIFT J0331.1+4355). The total effective exposure for GK Per is
9.5\,Ms. The analysis of the AAVSO light curve of the source 
%presented in Fig.~\ref{fig:swiftlc} 
shows that three outbursts occurred during the missions
lifetime and the averaged spectrum includes these intervals. On the other hand,
total duration of outbursts (1.4\,Ms) and average flux are comparatively low at
$\sim 4.5\times 10^{-11}$\,erg\,cm$^{-2}$\,s$^{-1}$ (in the 20-80 keV energy range),
so the quiescent emission is still likely to dominate the average spectrum.

To verify whether this is indeed the case, we also use the data from the INGEGRAL
IBIS \citep{IBIS} instrument including
all public observations within 12$^\circ$ from the source and excluding the
outburst periods as determined from the BAT light curve (in particular, we
exclude intervals when the \emph{Swift}/BAT flux is greater than 5 mCrab). These
results in a total of 1885 INTEGRAL pointings and an effective exposure of
$\sim0.5$\,Ms. GK Per is not detected in individual pointings, so we
extract the spectrum from the mosaic images obtained using all observations as
recommended in the INTEGRAL data reduction
guide\footnote{http://www.isdc.unige.ch/integral/download/osa/doc} for faint
sources. For data reduction we used the offline software analysis package OSA
10.1 and associated calibration files. The resulting spectrum has factor of two
lower flux than the mission long \emph{Swift}/BAT spectrum (i.e.
$2.5\times10^{-11}$\,erg\,cm$^{-2}$\,s$^{-1}$), but otherwise there is no
statistically significant difference in spectral shape. In particular, when
fitted with optically thin bremsstrahlung models, the best-fit temperature is
20.1 keV in both cases. Therefore we assume that both ISGRI and BAT spectra are
representative of source properties in quiescence and simultaneously fit both
with a free cross-normalization factor to account for the flux difference and
instrumental discrepancies in absolute flux calibration.

\section{Results}

\subsection{EX Hya} \label{subsect EX Hya} 
\begin{figure}
\centering
\includegraphics[angle=-0,scale=0.9]{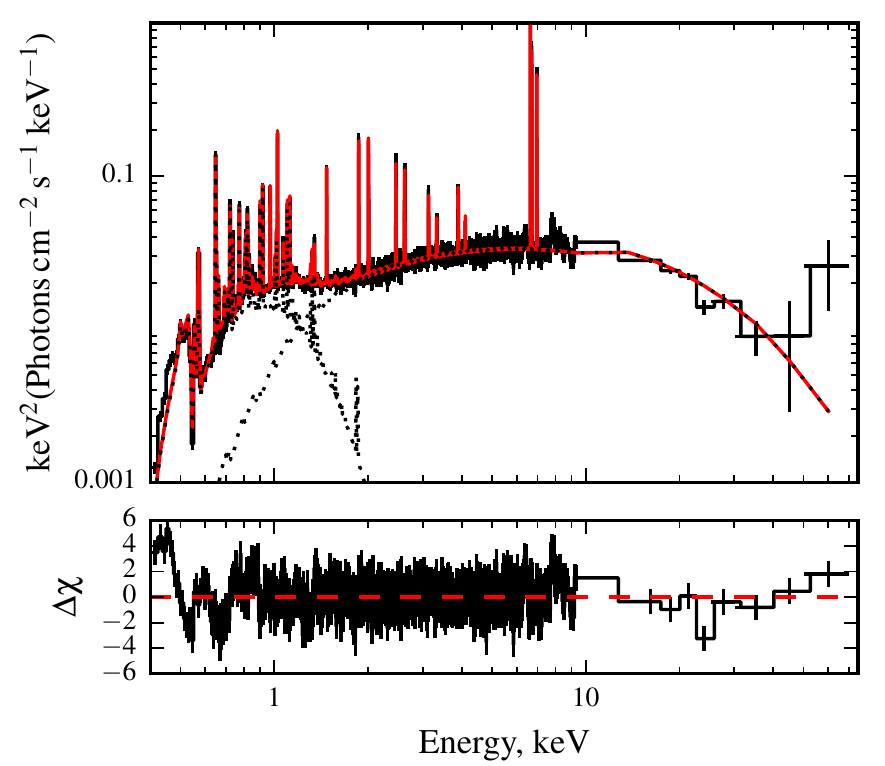}
\caption{\label{ex_sp}
\emph{Suzaku} spectrum of EX Hya fitted with the PSR model spectrum with $M/M_\odot = 0.79$
and $R_{\rm m}/R = 2$. 
%\textcolor{red}{\emph{in the Suzaku cookbook or somewhere else
%I remember that they suggest to use PIN data above aboute 15 keV.
%In this way you can remove the bin that produces the bad residual at about 15 keV...}}
}
\end{figure}

The broadband spectrum presented in Fig.~\ref{ex_sp} above
$\sim2$\,keV is well described with the absorbed PSR model with the WD mass
fixed to 0.79 $M_\odot$ as found from the binary motion \citep{BR:08}. At lower energies
there is a soft excess commonly observed in intermediate polars \citep{EH:07},
which can be accounted for either with a blackbody or APEC with kT $\sim$0.2
keV. Multiple narrow emission lines are known to be present in the soft band as
well \citep{Luna10} and some are apparent also in the residuals of the XIS
spectrum. To account for them we added several gaussians with zero widths and
energies fixed to values reported by \cite{Luna10} based on the high resolution
Chandra spectra of the source. We also included a cross-normalization constant
fixed to 1.18 %\textcolor{red}{\emph{I remember 1.16 from the Suzaku memo}} 
for the PIN spectrum (as suggested in data reduction guide for
PIN-nominal pointing). Other relevant parameters obtained from the fit are
$R_{\rm m}/R = 2.06^{+0.05}_{-0.04}$ and $N_{\rm H} = 4.1^{+0.1}_{-0.7}\times
10^{21}$\,cm$^{-2}$.
%The best fit model parameters are presented in
%Table.~\ref{tab:ex_hy_spepars}. 

The hard X-ray continuum ($\gtrsim 3$\,keV)
is adequately described by the PSR model.
The spectra residuals below $\sim 2$\,keV are caused by
unmodelled line and continuum emission.
We conclude, therefore, that the proposed model well
describes the X-ray continuum of the source.

The X-ray continuum is mostly sensitive to WD parameters in
the hard energy range, hence, the soft part of the spectrum does
not actually help to eliminate the model degeneracy between WD mass and
magnetosphere radius. In fact, more detailed analysis similar to that by
\cite{Luna10} would be required to derive additional constraints on WD
parameters from the soft X-ray spectra. Therefore, due to the strong absorption
and lack of broadband quiescent observations, we only considered the spectrum above
20\,keV for GK~Per (see below). Hereafter we ignore XIS data to estimate
the  mass of the WD in EX Hya. This approach
will allow us to verify whether the proposed method can
provide adequate results using hard X-ray data alone.

We used the combined light-curve with time resolution of
8\,s from the three XIS units active during the observation to obtain the power
spectrum of the source. As shown in Fig.~\ref{fig1b}, the power spectrum has a
break at $\nu_{\rm b} = 0.021 \pm 0.006$\,Hz.

%Under the same assumptions as discussed above for GK~Per, this translates to a strip in R/M diagram
%as shown in Fig.~\ref{fig:ex}

%The total \emph{Suzaku} spectrum of EX Hya, which includes the both spectra obtained by \emph{XIS} and \emph{XHD/PIN} instruments,
%is well fitted by computed PSR model spectra with the fixed  WD mass, $M=0.79 M_\odot$, see Fig.\,\ref{ex_sp}. Other parameters,
%such as $R_{\rm m} = 2.06^{+0.05}_{-0.04} R$ and  $N_{\rm H} = 4.1^{+0.1}_{-0.7}\cdot 10^{21}$\,cm$^{-2}$ were
%obtained by fitting procedure. To obtain a good fit of \emph{XIS} spectrum nine separate spectral lines described by Gauss emission lines   
%and additional \emph{APEC} component with the temperature $\approx$ 0.18 keV describing a well known soft excess in this source \citep[see, e.g.][]{EH:07}were 
%added. The resulting $\chi_{\rm d.o.f.}$ is $\approx$\,1.23.

\begin{figure}
\centering
\includegraphics[angle=0,scale=0.9]{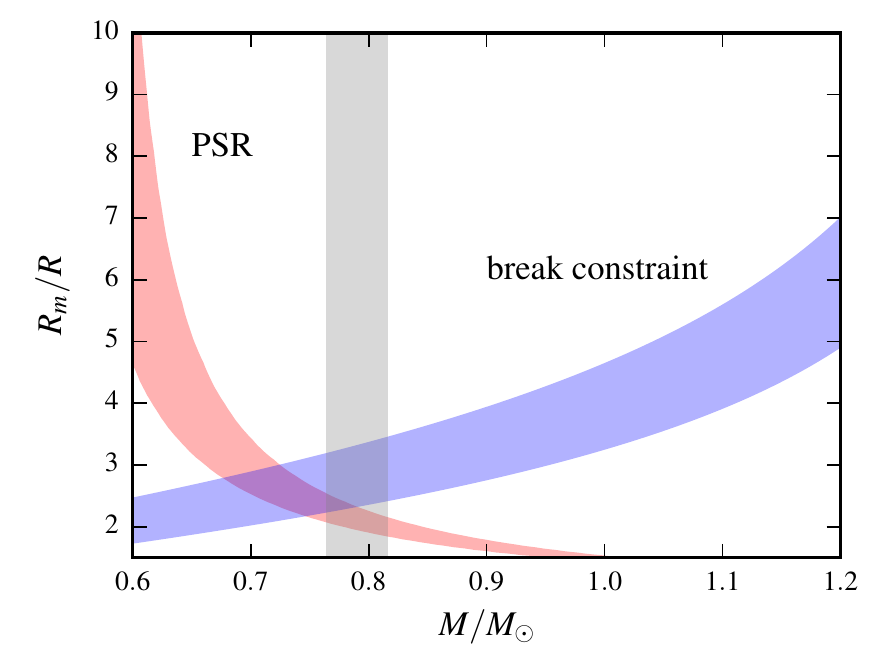}
\caption{\label{ex_cnt}
Strips in the $m - r_{\rm m}$ plane obtained using spectral fitting  and 
break frequency in the power spectrum of EX Hya. The vertical strip corresponds
to the WD mass known from the optical observations.
}
\end{figure}

%\textcolor{red}{\emph{Somewhere here you should put a reference to Fig. \ref{ex_cnt}}}
Next, we perform the two-parameter fitting of PIN data using model PSR spectra.
The resulting strip, which corresponds to the best fit including formal errors,
is shown in the $m - r_{\rm m}$ plane (see Fig.\,\ref{ex_cnt}). The strip, which corresponds to the
break frequency $\nu_{\rm b}$ with the uncertainties  (see Fig.\,\ref{fig1b}) is
also shown. The crossing of the two regions allows us to find the best fit
parameters of EX Hyd, $M /M_\odot = 0.73 \pm 0.06 $, and $R_{\rm m}/R = 2.6\pm0.4$. 
The obtained parameters coincide within errors with the values obtained
using other methods (see Sect. \ref{sect EX Hya}). 
Therefore, we conclude that the suggested method gives
reliable results for EX Hya, and can be used for other
intermediate polars.

\subsection{GK Per}
\subsubsection{Outburst data}
\begin{figure}
\centering
\includegraphics[angle=0,scale=1.0]{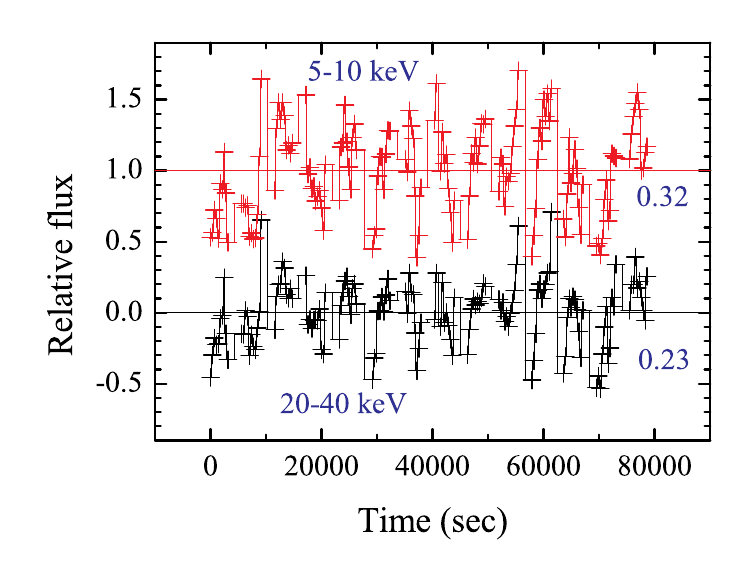}
\caption{\label{fig6}
Relative light curves of GK Per in two spectral bands, 5-10 keV and 20-40 keV
and time resolution of 351\,s.
The fluxes in two bands are correlated, but the RMS of the light curve in the soft band
is higher (0.32) in comparison with the RMS of the light curve in hard band (0.23).  
}
\end{figure}

Figure \ref{fig6} shows the \emph{NuSTAR} background subtracted light curves of the source
in two energy bands. The source exhibits correlated
variability in both energy bands on timescales of 5-10\,ks which resembles the
quasi-periodical oscillations (QPOs) reported earlier \citep{Watson.etal:85,
Ishida.etal:96}. However, we do not formally detect the QPOs directly in power
spectrum due to the data gaps in the \emph{NuSTAR} observation which occur on the
same timescale.

\begin{figure}
\centering
\includegraphics[angle=0,scale=0.9]{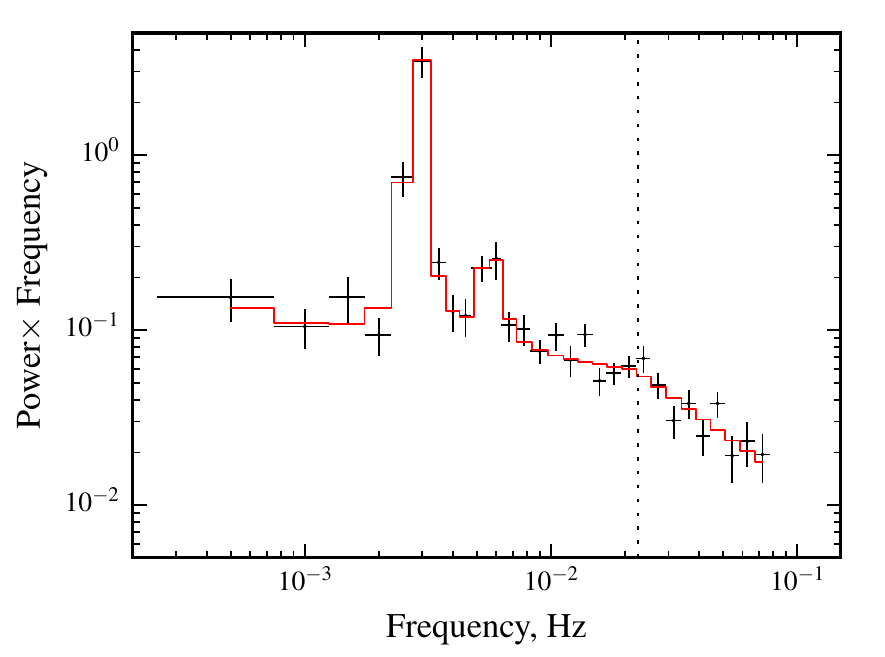}
\caption{\label{fig7}
Power spectrum of the X-ray light curve of GK Per observed by \emph{NuSTAR}. The found break frequency $\nu_{\rm b} = 0.0225\pm 0.004$
is marked with the vertical dotted line. The peaks are corresponded to the spin frequency and its first harmonic. 
}
\end{figure}

The most prominent features in the power spectrum of the 3-80\,keV lighcurve
are the two peaks associated with the spin frequency and the fist harmonic, as
well as the break at $\nu_{\rm br}=0.0225\pm0.004$ 
%\textcolor{red}{\emph{in the caption of the figure it is 0.0224}} 
(Fig~\ref{fig7}). 

\begin{figure}
\centering
\includegraphics[angle=0,scale=0.9]{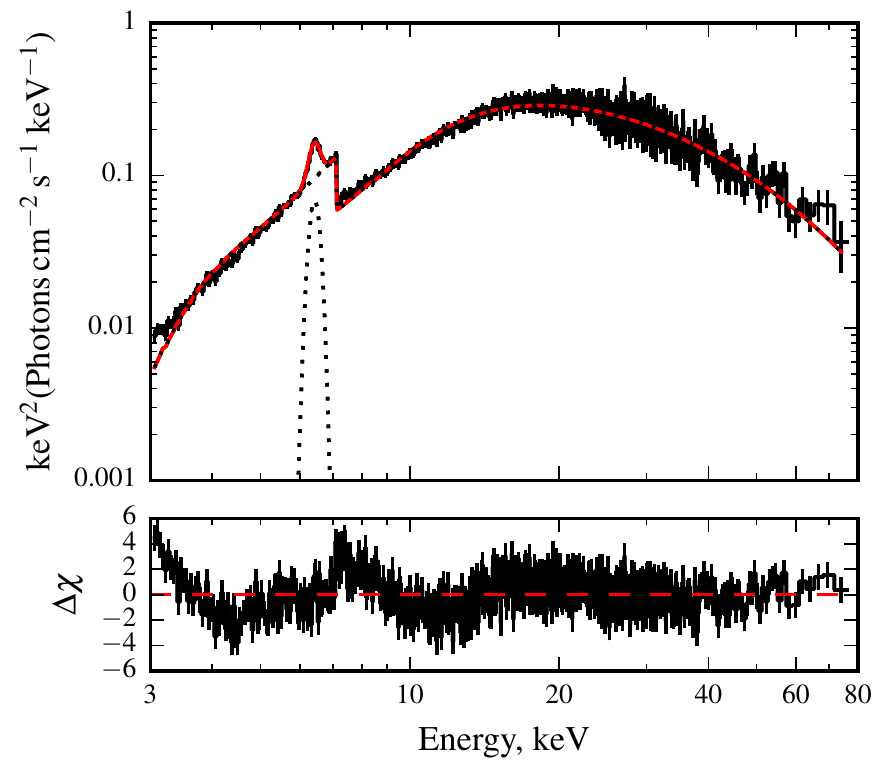}
\caption{\label{gk_sp}
Unfolded \emph{NuSTAR} spectra of GK Per. The model spectrum was computed for
fixed $M/M_\odot=0.86$. Other parameters found are $R_{\rm m}/R \approx 2.3$ and
 $N_{\rm H,1} \approx 1.7 \cdot 10^{23}$\,cm$^{-2}$ . The obtained partial covering parameters are
$N_{\rm H,2} \approx 1.24\cdot 10^{24}$\,cm$^{-2}$ and covering fraction 
$C_{\rm F} \approx 0.82$, and $\chi^2_{\rm dof} \approx 1.75$.
 }
\end{figure}

QPOs were previously associated with obscuration of the
emission from the white dwarf by bulges in the inner disc \citep{HL:94}.
% This hypothesis
%is consistent with significantly increased absorbtion during the outbursts
%(ref1) when the accretion disk is expected to push the magnetosphere inwards
%thus making such obscuration more probable. 
We note that the amplitude of variability in the \emph{NuSTAR} data decreases with
increasing energy (see Fig.\,\ref{fig6}), which is also consistent with the hypothesis
that the flux variability is largely driven by changes in the absorption column. To
verify this hypothesis, we fitted the averaged
%extracted the
spectrum 
%from time intervals when the pulse-averaged count-rate in 3-80 keV band
%exceeded and was below the average value of 8 (16 ????)\,counts/s separately. We fited
presented in Fig.~\ref{gk_sp} with the PSR model modified by partial covering
absorber \citep[see, e.g., discussion in][]{Ramsay:00}.
%with only covering fraction and
%normalization allowed to vary, whereas the WD mass (0.86 $M_\odot$), 
%the magnetospheric radius (2.8 $R$), and the absorbing column density
% ($N_{\rm H}\,=1.???\times  10^{24}$\,cm$^{-2}$) are fixed. 
First, it is important to emphasize that the higher statistical quality of
\emph{NuSTAR} data makes it clear that this simple model is not really adequate
to describe the broadband spectrum of the source below 20 keV. Nevertheless,
the spectrum is well described by the
model above $\sim 20$\,keV.
% and the only difference besides the average flux is the convering
%fraction which increased from $\sim0.8$ at higher flux to $\sim0.86$ at lower
%flux. Alternatively, the observed suppression of flux at low energies can be
%accounted for by increased absorbtion.

%\begin{figure}
%\centering
%\includegraphics[angle=0,scale=0.5]{pulse_profiles.pdf}
%%\includegraphics[angle=0,scale=1.]{fig1n.eps}
%\caption{\label{fig:pp}
%Normalized pulse profiles of GK Per in four energy bands. 
%}
%\end{figure}

We note also that the light curves folded with the spin period of the white dwarf
show no significant dependence on the energy 
with the pulsed fraction $PF \approx 45\%$  in all 
the energy bands. Therefore, the absorbing material is likely not located in
the immediate vicinity of the WD. 
In addition, rapid variation of source hardness with
time suggests that the absorption is variable on relatively short timescales
and that the partial covering absorber is just a useful approximation to
account for constantly changing absorption column 
which is likely caused by the
obscuration of the emission region by outer parts of the accretion disk.
%\textcolor{green}{REMOVE? Indeed, it is possible to progressively improve the fit by introducing
%additional absorbing columns. However, such approach is not unambiguous.} 
%\textcolor{red}{\emph{here it is not clear to me what do you mean with ``not unambiguous''.}}.

Moreover, the hard part of the spectrum is mostly sensitive to the WD
parameters whereas the largest fraction of photons are detected in 
the soft part which is dominated
by complex absorption (as expected for the derived absorption column
of $10^{23-24}$\,cm$^{-2}$). We note that while it is possible to describe the
broadband spectrum introducing several absorption columns, any ambiguity in modeling
of the soft part of the spectrum is likely to affect also the hard part, and
thus the derived parameters of the WD simply due to the fact that
statistically the soft part is much more important. Therefore, to avoid any
potential systematic effects associated with modeling of the soft part of the
spectrum we conservatively ignore data affected by the absorption below 20\,keV
to determine the parameters of the WD.

\begin{figure}
\centering
\includegraphics[angle=0,scale=0.42]{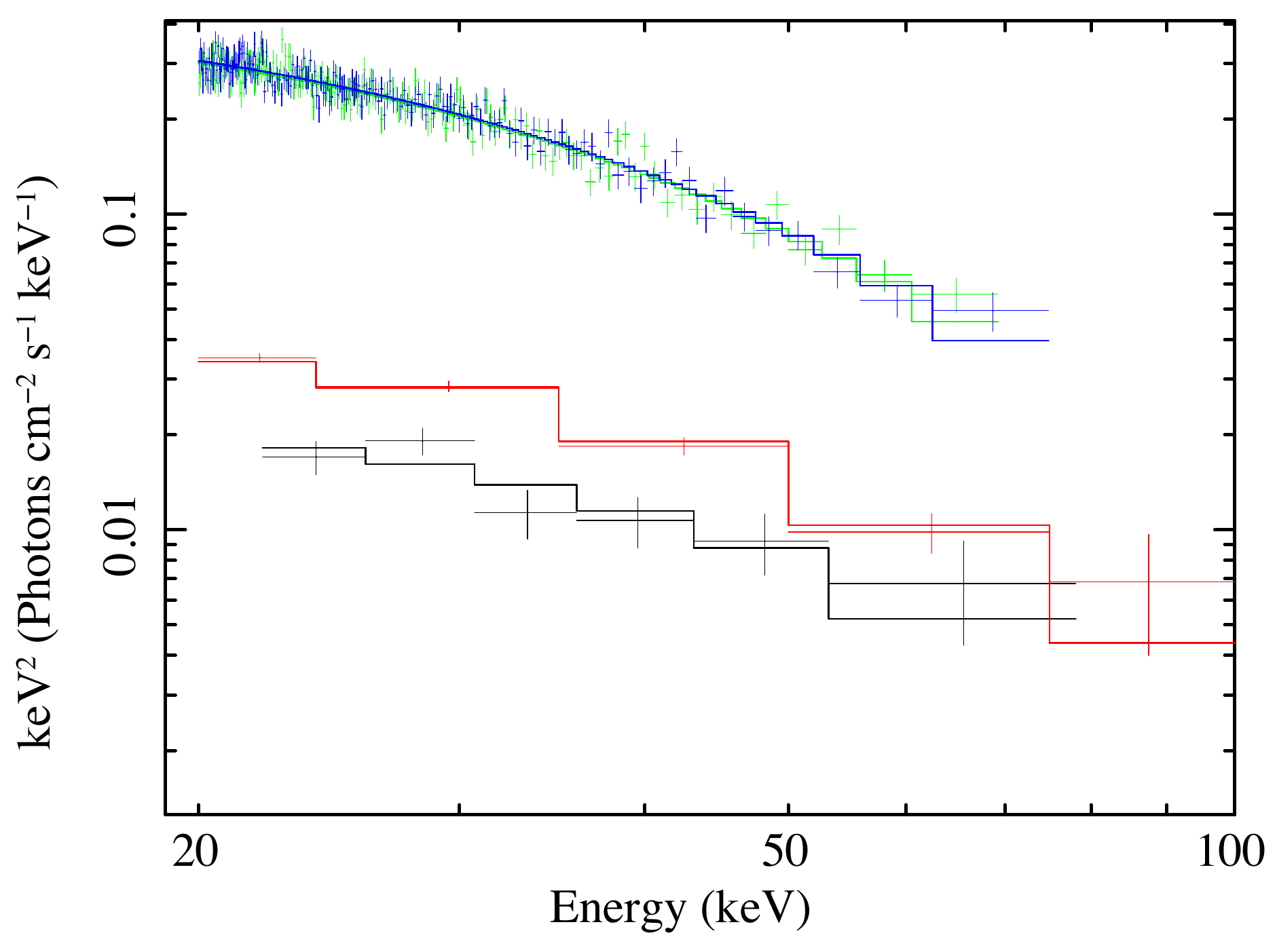}
\caption{\label{gk_all}
Observed spectra of GK Per used for two-parameter spectral fitting, from top downward: \emph{NuSTAR} spectrum above
20 keV, \emph{Swift}/BAT, and INTEGRAL spectra.
% \textcolor{red}{Here the right vertical axis is very thinwhen I print the paper.
}
\end{figure}

\begin{figure}
\centering
\includegraphics[angle=0,scale=0.9]{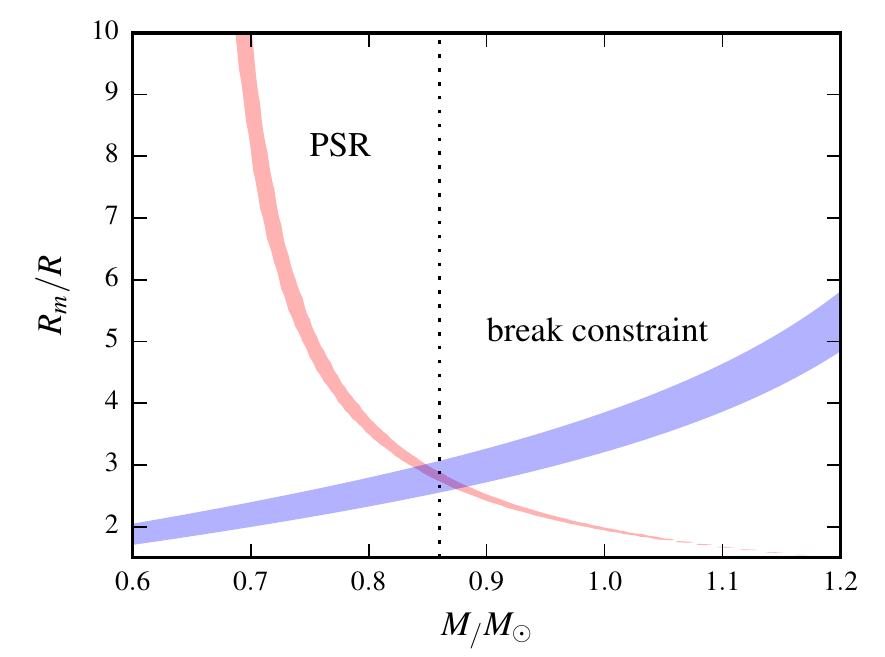}
\caption{\label{gk_nu_c}
Strips in the $m - r_{\rm m}$ plane obtained using spectral fitting  \emph{NuSTAR} spectrum
 and break frequency in the power spectrum of GK Per in outburst.
}
\end{figure}

The hard X-ray \emph{NuSTAR} spectrum used to estimate the WD parameters in GK Per is shown in Fig.\,\ref{gk_all}.
The resulting strips in the $m - r_{\rm m}$ plane are shown in Fig.\,\ref{gk_nu_c}. The intersection region yields
$M/ M_\odot= 0.86 \pm 0.02$ and $R_{\rm m}/R\, =\, 2.8\pm 0.2$. The value of the mass is consistent
with the values determined by other authors (see Sect.\,\ref{gkper}), 
but our measurement has much better accuracy.

\subsubsection{Magnetosphere size}

\begin{figure}
\centering
\includegraphics[angle=0,scale=0.9]{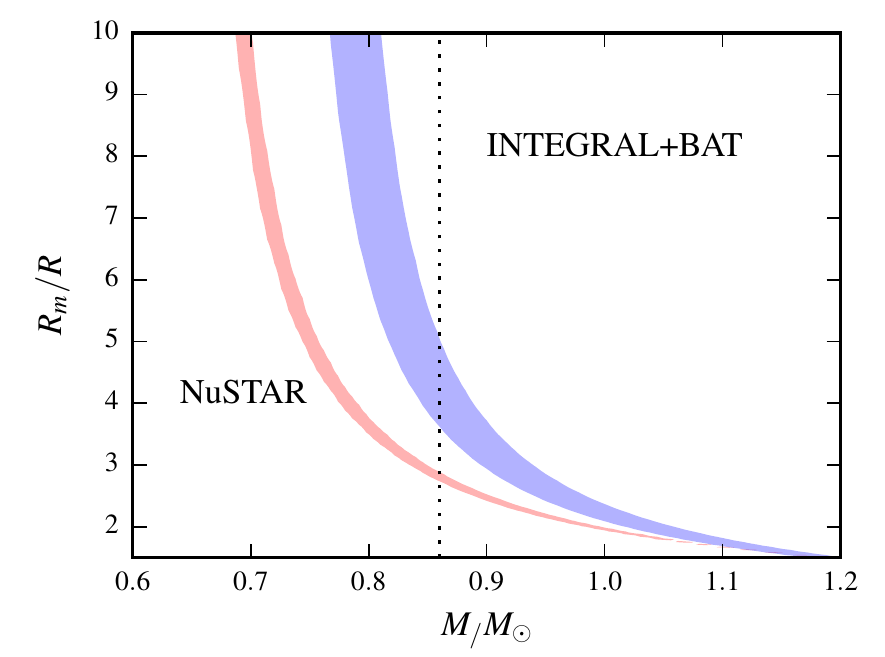}
\caption{\label{gk_all_c}
Strips in the $m - r_{\rm m}$ plane obtained using spectral fitting of GK Per 
in outburst (\emph{NuSTAR} spectrum) and quiescence (combined \emph{Swift}/BAT and INTEGRAL
spectrum).
}
\end{figure}

The hard X-ray luminosity of GK Per in quiescence is almost an order of
magnitude lower in comparison with the outburst (Fig.\,\ref{gk_all}). Fitting
the spectra using the bremsstrahlung model yields 
$kT_{\rm br} = 16.7\pm 0.2$\,keV
(\emph{NuSTAR} spectrum in the outburst), $kT_{\rm br} = 20.1\pm 0.8$\,keV 
(\emph{Swift}/BAT averaged spectrum), 
and $kT_{\rm br} = 21^{+4}_{-3}$\,keV (\emph{INTEGRAL} spectrum in the quiescence), 
i.e. when the luminosity decreases the magnetosphere size increases,
thus the temperature increases.
To quantify this effect and to obtain the observed dependence $m -
r_{\rm m}$ in quiescence we fitted the \emph{Swift}/BAT and \emph{INTEGRAL}
spectra simultaneously. The result is shown in Fig.\,\ref{gk_all_c}.

\begin{figure}
\centering
\includegraphics[angle=0,scale=1.1]{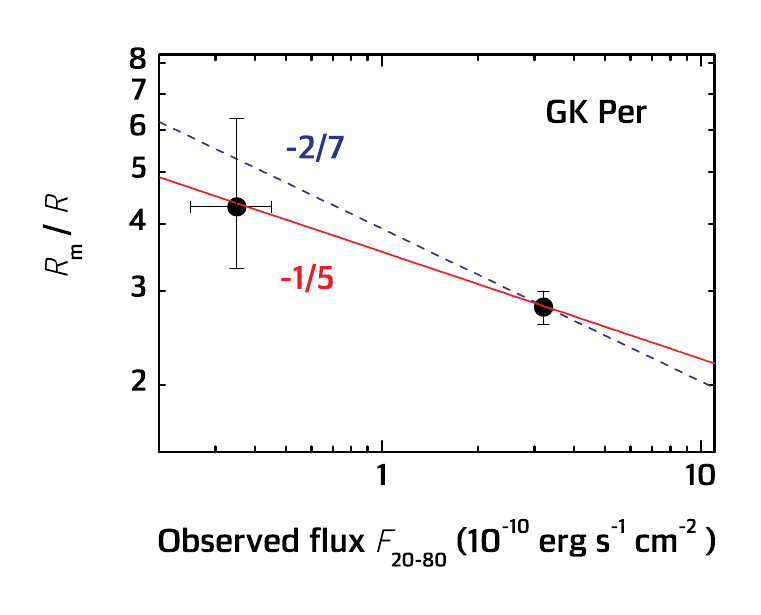}
\caption{\label{rm}
Dependence of the GK Per magnetospheric radius on the observed flux in the 20-80 keV band.
The best fit with the slope -1/5 is shown together with the slope -2/7 
inferred from the equation for the Alfven radius.
}
\end{figure}

It is clear the magnetospheric radius in GK Per indeed increases in quiescence,
and equals $4.3^{+2}_{-1}\, R$ for $M = 0.86$\,$M_\odot$. 
Therefore, we can investigate how the magnetospheric radius depends on the observed X-ray
flux and thus the accretion rate (Fig.\,\ref{rm}). Here we used the observed flux
of GK Per in the 20-80\,keV band $F_{20-80}$ as a tracer of the accretion rate. The
magnetosphere size is expected to scale as some power of luminosity or accretion rate.
% and power
%law fit to the two points \textcolor{blue}{on the plane $r_{\rm m} - F$} obtained above 
%\textcolor{red}{\emph{sorry, it is not clear where they are obtained,probably I should read it again carefully!}}
We evaluate the exponent $\beta$ in the
$r_{\rm m} \sim (F_{20-80})^{\,\beta}$ dependence to $-0.2^{+0.10}_{-0.15}$ using two obtained points on the
$r_{\rm m} - F_{20-80}$ plane. Formally,
the classical exponent in the equation for the Alfven radius $\beta_{\rm af} =
-2/7$ is well consistent with the obtained value (which is rather uncertain due
to the low statistics in quiescence). On the other hand, it is interesting to note
that similarly to us, \citet{KR:13} found  a somewhat lower value than the classical one
using 3D MHD simulations for small magnetospheres, with $r_{\rm m}$ in the
range 2.5-5: 
\be \label{rmkr} 
\frac{R_{\rm m}}{R} \approx 1.06\,\left(\frac{\mu_{\rm B}^4}{\dot M^2\,GM\,R^7}\right)^{1/10}, 
\ee 
where $\mu_{\rm B}\approx BR^3$ is the magnetic moment of the WD, and $B$ is the magnetic field
strength on its surface. \citet{KR:13} explained this result with
compressibility of the magnetosphere and remarked that this effect is
less significant for larger magnetospheres. Currently our estimate is rather
uncertain due to low statistics in quiescence and not constraining the theory.
However, the conclusions by \cite{KR:13} might become testable once better spectra
of GK~Per at low fluxes are available.

%We evaluate magnetic field strength on the WD surfaces in EX Hya and GK Per using Eq.(\ref{rmkr}) and the expression for the accreting X-ray luminosity
%\be
%         L_{\rm X} = \frac{GM\dot M}{R}\,\left(1-\frac{R}{R_{\rm m}}\right).
%\ee  
%X-ray luminosity of EX Hya is about 2.6 $\times  10^{32}$\,erg s$^{-1}$ and it gives $B \approx 10^4$\,G. GK Per is much luminous with 
%$L_{\rm X} \approx ??\times  10^{34}$\,erg s$^{-1}$, and, therefore, WD is magnetized stronger in this CV with $B \approx 10^5$\,G.

\section{Conclusions}
We suggested a new method for simultaneous determination of the white dwarf
mass and the magnetospheric radius in intermediate polars. The method is based
on two independent measurements of the degenerate $M-R_{\rm m}$ dependence
using the observed break frequency in the power spectrum 
\cite[see][]{Revnivtsev.etal:09, Revnivtsev.etal:11}, and
the fitting of the hard X-ray spectrum with the newly developed PSR model
which takes into account the finite acceleration height of the accretion flow.
The two measurements lead to two intersecting regions 
in the  $M/M_\odot -R_{\rm m}/R$ plane which allow to estimate the white dwarf mass
and the relative magnetospheric radius.

For the spectral fitting procedure, we 
computed an extensive grid of two-parameter models for hard X-ray spectra of
post-shock regions on WD surfaces. We assumed
quasi-dipole geometry of the PSR, fixed accretion rate ($\dot M = 10^{16}$\,g
s$^{-1}$) and a polar region with fixed relative footprint $f = 5\times 10^{-4}$
of the WD surface. The cyclotron cooling and difference in temperatures of ion
and electron plasmas are currently not taken into account. The WD mass
range covered is 0.3 -1.4 $M_\odot$ (with steps of 0.02 $M_\odot$).
The second parameter of the model, the relative magnetospheric radius $r_{\rm m} =R_{\rm
m}/R$, changes from 1.5 to 60 with steps proportional to
$(1-r_{\rm m}^{-1})$ (we also included a model with $r_{\rm m} = 1000$ to ease
comparison with previously published results, i.e. the grid includes 41 values of
$r_{\rm m}$). 
The model is implemented in the XSPEC package as an additive table model
and accessible to the scientific community.

The method was tested using the well studied intermediate polar EX Hya. We
obtained  the WD mass of (0.73$\pm 0.06) M_\odot$ and magnetospheric radius $R_{\rm m}/R =
2.6 \pm 0.4$, which are fully consistent with the known WD mass and the 
magnetosphere size expected for this source. Subsequently we applied the method to another 
intermediate polar GK~Per, which is also a dwarf nova.
Large changes of flux during the outburst in GK~Per allow not only to estimate
the WD mass and the relative magnetosphere size, but to also to investigate
the magnetosphere size changes with luminosity. 

Using the \emph{NuSTAR} observation of GK~Per during an outburst at the flux
level of $F_{\rm 20-80} = 3.2\times 10^{-10}$\,erg s$^{-1}$\,cm$^{-2}$ in the
20-80\,keV range, we estimate the WD mass to $M/ M_\odot= 0.86 \pm 0.02$ and
$R_{\rm m}/R = 2.8\pm 0.2$. The fit to the combined \emph{Swift}/BAT and INTEGRAL
spectra of GK Per in quiescence gives $R_{\rm m}/R = 4.3^{+2}_{-1}$ at fixed
$M$ and $F_{\rm 20-80} = 3.5 \times 10^{-11}$\,erg s$^{-1}$\,cm$^{-2}$. The derived
$R_{\rm m}/R - (F_{\rm 20-80})\,^\beta$ dependence with $\beta =
-0.2^{+0.10}_{-0.15}$ is consistent with the classical dependence for Alfven
radius and with the results obtained by \citet{KR:13} for small magnetospheres
from MHD simulations. We note that it could be possible to test the
predictions by these authors once the quiescent hard spectra of GK~Per with
better statistical quality are available.

\begin{acknowledgements} 
This work was  made use of data from the \emph{NuSTAR}
mission, a project
led  by  the  California  Institute  of  Technology,  managed  by  the  Jet  Propulsion
Laboratory, and funded by the National Aeronautics and Space Administration.
This paper is also based on data from observations with INTEGRAL,
an ESA project with instruments and science data centre funded by ESA
member states (especially the PI countries: Denmark, France, Germany,
Italy, Spain, and Switzerland), Czech Republic and Poland,
and with the participation of Russia and the USA.
This research has made use of data obtained from the \emph{Suzaku} satellite,
a collaborative mission between the space agencies of Japan (JAXA)
and the USA (NASA). 
V.S. thanks Deutsche
Forschungsgemeinschaft (DFG) for financial support (grant 
WE 1312/48-1) .
V.D. and L.D. acknowledges support by the 
Bundesministerium f\"ur Wirtschaft
und Technologie and the Deutsches Zentrum f\"ur Luft und Raumfahrt through
the grant FKZ 50 OG 1602.

\end{acknowledgements}

\bibliographystyle{aa}
\bibliography{ip}

\appendix
\label{app}
\section{Compare with the previous paper}

\begin{figure}
\centering
\includegraphics[angle=0,scale=1.]{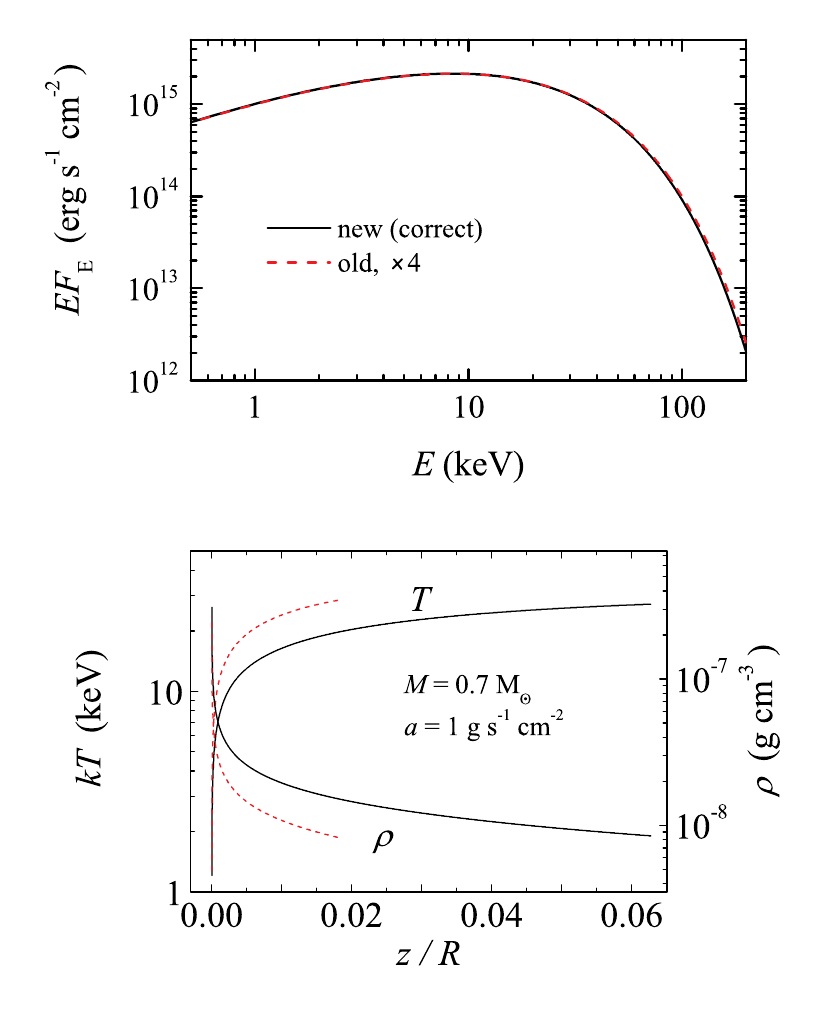}
\caption{\label{fig15}
Old and new spectra (top panel) as well as temperature and density stratifications (bottom panel).
}
\end{figure}

As mentioned by \cite{HI:14a}, in the paper \citet{SRR:05} the total number density $n$ was used instead of both the ion number density $n_{\rm i}$ 
and the electron number density $n_{\rm e}$.
That replacement increased the cooling rate in the considered accretion column models approximately four times making a column model four times shorter
(see Fig.\,\ref{fig15}, bottom panel). This led to a different emergent spectrum normalization (four times smaller).
Fortunately, the shapes of the computed spectra were not affected  
(see Fig.\,\ref{fig15}, top panel).  Consequently, the WD masses derived in the \citet{SRR:05}  remain correct, because they are determined by the shape of the
spectrum only. We note, that this error was found just after the publication, and the spectrum grid used by \citet{Betal:09} was recomputed
with a corrected version of the code. Models computed with Compton scattering taken into consideration \citep{SPFW:08}
 were also computed using the correct cooling rate.  

%\begin{thebibliography}{24}
%\expandafter\ifx\csname natexlab\endcsname\relax\def\natexlab#1{#1}\fi

%\end{thebibliography}

%\clearpage

\end{document}